\pdfoutput=1

\documentclass[a4paper, 10pt, conference]{IEEEtran} 
\usepackage{times}

\usepackage{graphicx}
\usepackage{epstopdf}
\usepackage{multirow, makecell}
\usepackage{balance}
\usepackage{enumitem}
\usepackage{url}
\usepackage{amsmath} 

\newcommand\blfootnote[1]{%
  \begingroup
  \renewcommand\thefootnote{}\footnote{#1}%
  \addtocounter{footnote}{-1}%
  \endgroup
}

\renewcommand{\footnoterule}{%
  \kern -3pt
  \hrule width \columnwidth height 1pt
  \kern 2pt
}

\usepackage{color}
\makeatletter
 \let\old@ps@headings\ps@headings
 \let\old@ps@IEEEtitlepagestyle\ps@IEEEtitlepagestyle
 \def\confheader#1{%
 \def\ps@headings{%
 \old@ps@headings%
 \def\@oddhead{\strut\hfill#1\hfill\strut}%
 \def\@evenhead{\strut\hfill#1\hfill\strut}%
 }%
 \def\ps@IEEEtitlepagestyle{%
 \old@ps@IEEEtitlepagestyle%
 \def\@oddhead{\strut\hfill#1\hfill\strut}%
 \def\@evenhead{\strut\hfill#1\hfill\strut}%
 }%
 \ps@headings%
 }
 \makeatother
 
\makeatletter
\def\ps@IEEEtitlepagestyle{
  \def\@oddfoot{\mycopyrightnotice}
  \def\@evenfoot{}
}
\def\mycopyrightnotice{
  {\footnotesize
  \begin{minipage}{\textwidth}
  \centering
  Author's accepted manuscript version. Accepted for publication in IEEE ICC 2021, IoT and Sensor Networks Symposium
  
  The current author's accepted manuscript version is released for the purpose of meeting public availability requirements. 
   
  Please refer to the published version, for citing this work or for other possible usage. 
  
“\copyright 2021 IEEE.  Personal use of this material is permitted.  Permission from IEEE must be obtained for all other uses, in any current or future media, including reprinting/republishing this material for advertising or promotional purposes, creating new collective works, for resale or redistribution to servers or lists, or reuse of any copyrighted component of this work in other works.”
  
  \end{minipage}
  }
}
 
\begin{document}

%
%
%


\title{\textbf{Analysing the Data-Driven Approach of Dynamically Estimating Positioning Accuracy}\\}

\author{\IEEEauthorblockN{~\large Grigorios G. Anagnostopoulos, Alexandros Kalousis}
\IEEEauthorblockA{Geneva School of Business Administration, HES-SO
Geneva, Switzerland\\
Email:  $\left\{ grigorios.anagnostopoulos, alexandros.kalousis \right\}$@hesge.ch}
}

\maketitle

\begin{abstract}
The primary expectation from positioning systems is for them to provide the users with reliable estimates of their position. An additional piece of information that can greatly help the users utilize position estimates is the level of uncertainty that a positioning system assigns to the position estimate it produced. The concept of dynamically estimating the accuracy of position estimates of fingerprinting positioning systems has been sporadically discussed over the last decade in the literature of the field, where mainly handcrafted rules based on domain knowledge have been proposed. The emergence of IoT devices and the proliferation of data from Low Power Wide Area Networks (LPWANs) have facilitated the conceptualization of data-driven methods of determining the estimated certainty over position estimates. In this work, we analyze the data-driven approach of determining the Dynamic Accuracy Estimation (DAE), considering it in the broader context of a positioning system. More specifically, with the use of a public LoRaWAN dataset, the current work analyses: the repartition of the available training set between the tasks of determining the location estimates and the DAE, the concept of selecting a subset of the most reliable estimates, and the impact that the spatial distribution of the data has to the accuracy of the DAE. 
The work provides a wide overview of the data-driven approach of DAE determination in the context of the overall design of a positioning system.
\end{abstract}

\renewcommand\IEEEkeywordsname{Keywords}

\IEEEpeerreviewmaketitle

\blfootnote{
\\
Code: \url{https://doi.org/10.5281/zenodo.4099079}   \\
Data: \url{https://doi.org/10.5281/zenodo.4117818} 

}

%


\section{Introduction} \label{sec:Introduction}

Interest in the utilization of IoT networks has been increasing over recent years. Moreover, the massive amounts of data that the IoT connectivity produces, invite intelligent ways of utilizing this massive data volume. In this direction, the emergence of Low Power Wide Area Networks (LPWANs) has offered the grounds for energy efficient localization applications. The data-driven methods of localization, called \textit{fingerprinting} methods, which are a common solution for indoor positioning systems, have recently been applied in outdoor IoT settings~\cite{Sigfox_Dataset, Janssen_Sigfox, Anagnostopoulos2019LoRa,Janssen_2020_Benchmarking, Anagnostopoulos2019_IPIN,Hybrid_LoRaWAN_2021}. Apart from the production of location estimates, the positioning systems may also produce a \textit{Dynamic Accuracy Estimation} (DAE), which quantifies the certainty of the system over its produced location estimate~\cite{Lemic_regression_based_2019, ANN_Lemic_2020}. The broad public has been familiarized with this concept, since the widely used GPS, along with other Global Navigation Satellite Systems (GNSSs), often offer the users a point on a map indicating their estimated location, along with a circle centered at the location estimate, with a radius of a size that is proportionate to estimated potential error of the provided location estimate. 

The DAE can be perceived as the claimed accuracy of a position estimate, according to the positioning system that produced said estimate. As the name states, DAE is produced dynamically, in an on-line manner, along with the production of the location estimates, without having access to the ground truth location. Thus, it concerns an estimation of the accuracy and not a measurement of the accuracy.

In this work, we elaborate on the concept of Dynamic Accuracy Estimation, in a data-driven approach. In order to calculate the DAE in a data-driven manner, a model is trained, receiving the same feature set as the model that produces the position estimates, while the targets of the two models are different. Instead of training the model having the ground truth location as target, as the positioning model does, the target this time is the distance between the ground truth location and the location estimate produced by the positioning model. Thus, this second model learns to predict the amount of the localization error of the location estimates of the first model. In this way, when the system processes a new reception of signals at an unknown location, the two models will be able to produce: (i) a location estimate, provided by the first (positioning) model, and (ii) an estimate of the localization error, estimated by the second (DAE related) model.

The potential of the use of DAE depends on the particularities of the business case in which a positioning system is meant to be used. The main idea behind the use of DAE is that it can be used either to offer the user a confidence level accompanying each position estimate, or to facilitate a higher-level system in taking related decisions. A first example of such decision making could be the use of DAE in hand-off algorithms, which could use the DAE as a reliability measure for switching among different technologies. Another example could be the option to present to the user only a percentage of the most accurate position estimates, using DAE as the accuracy indicator, should that suit a particular business case. For instance, in business cases where a very frequent update of the location estimates is not required, the system could select to return only a subset of the produced position estimates based on their accuracy, as estimated by the DAE. This concept is exemplified in the tests presented in this work. 
Alternatively, simply offering a confidence level on the position estimates that are produced could facilitate other modules of an IoT system in taking relevant decisions about the way they will utilize the location estimates.
 
The rest of this paper is organized as follows.
In Section~\ref{sec:Related}, the related work is discussed. Section~\ref{sec:Dataset} presents in detail the dataset used in this work, while Section~\ref{sec:Methodology} presents the methodology and defines the terminology used. Subsequently, Section~\ref{sec:Results} presents the experiments of this work and discusses the obtained results. Finally, conclusions drawn and ideas for future work are presented in Section~\ref{sec:Conclusions}.
	
\section{Related Work} \label{sec:Related}

The concept of on-line error estimation of the location estimates that a positioning system produces has been sporadically studied in the literature of the field, over more than a decade. This concept of error prediction is far from constituting a usually advertised feature of indoor positioning systems and relevant publications, or a subject of comparison in the competitions of the field~\cite{Torres_Sospedra_competition_2017}  which mainly compare the accuracy of the location predictions. Nevertheless, a framework of evaluating~\cite{GpmLab_2016} and tuning~\cite{GpmStudio_2016,GpmStudio_MO_2017} positioning systems, proposed in 2016, lists the DAE as one of the metrics that the framework can evaluate and  tune. 
Different approaches have been proposed regarding the way in which the certainty over a produced location estimate can be estimated. The categorical division of these approaches can be described by two main types of methods.

In the first category of the \textit{rule-based methods}, which monopolized the literature of the field for a long period of time, the aim has been to hand-craft analytic or heuristic methods of estimating the quality of location estimates. A variety of rule-based methods has been proposed, such as using: the average geographic distance between the nearest neighbors~\cite{Lemelson_2009}, improved by factoring in as a weight the proximity of the nearest neighbors  and also introducing the location estimate~\cite{Marcus_SMARTPOS_2013}, or by introducing a weighted average of likelihoods instead of a simple average of Euclidean distances~\cite{Li_Toward_2019}; a mechanism to calculate a Dilution-of-Precision-like value~\cite{Moghtadaiee_Accuracy_indicator_2012}; 
 the geographic distance from the furthest neighbor to the location estimate~\cite{Zou_Estimation_2014}; the spatial distribution of the latest position estimates~\cite{Elbakly_CONE_2016,Syed_Khandker_2019}; and an offline method based on the Cramer-Rao Lower Bound ratio~\cite{Nikitin_laousias_2017}.

The notion of the estimating the error of position estimates has been further utilized in publications dealing with handoff algorithms, switching between indoor and outdoor environments, or among different technologies. Lin et al.~\cite{Energy_Accuracy_Lin_2010} used a constant typical error value for each technology of their system (GPS, Bluetooth, WiFi) in a system logic that selects the most energy-efficient technology that satisfies a user-defined accuracy requirement threshold. Zou et al.~\cite{Zou_handoff_2013} utilized the Euclidean distance between the received RSS vector by the mobile device and its nearest neighbor from the training set as an indicator of whether the device in the service area of the used technology. This reliability indicator is then used in the proposed dual-threshold handoff algorithm. Similarly, Anagnostopoulos et al.~\cite{Anagnostopoulos_IPIN_2015}, proposed a hand-off algorithm which compares the claimed accuracy of two technologies in order to hand-off the control between the two of them. In~\cite{Anagnostopoulos_IPIN_2015}, a ranging Bluetooth positioning method is used and the estimated distance from the $3rd$ closest detected beacon is used as an empirically drawn metric of estimating the DAE. The focus in both~\cite{Zou_handoff_2013} and~\cite{Anagnostopoulos_IPIN_2015}  is on distinguishing whether the mobile device is inside the service area of a certain technology or not. These works do not explicitly evaluate the performance of the DAE per se, but only in the way that the DAE estimates allow the evaluated handoff algorithms to perform as expected.

On the other hand, the \textit{data-driven methods}, which constitute the second category that recently emerged, aim at learning to predict the quality of location estimates based on a dataset used to train a machine learning model. The current work focuses on this second, data-driven approach. A surprisingly early proposal of a data-driven method of DAE determination dates back to 2007, when Dearman et al.~\cite{Dearman_exploration_2007} proposed a multiple linear regression approach. That work uses designer-defined features based on the RSS fingerprints to train a model, which predicts the localization error (in meters) of a location estimate produced by a new signal reception. 

A pair of recent works, by Lemic et al.~\cite{Lemic_regression_based_2019} and Lemic and Famaey~\cite{ANN_Lemic_2020} has revived the data-driven approach. The availability of large datasets of IoT-based, LPWAN networks, such as the ones presented by Aernouts et al.~\cite{Sigfox_Dataset}, oreoares the ground for such approaches. In 2019, Lemic et al.~\cite{Lemic_regression_based_2019} presented an extensive analysis of the performance of several well-known regression algorithms (linear regression with regularizers, kNN, SVN, Random Forests) on predicting the positioning error of location estimates. 
In a natural continuation of that work~\cite{Lemic_regression_based_2019}, Lemic and Famaey~\cite{ANN_Lemic_2020} investigated the capabilities of Neural Networks (NN) in addressing the same regression problem. By utilizing the LPWAN datasets of~\cite{Sigfox_Dataset}, the authors show that the NN approach outperforms kNN, which was the best performing method of~\cite{Lemic_regression_based_2019}. 

In the current work, we further analyze the data-driven approach of determining the DAE, considering it in the context of a broader positioning system. In this context, we focus on points such as the repartition of the available training data between the models producing location estimates and DAE estimates, the location estimate selection process based on the DAE, as well as the impact of the spatial distribution of the data on the evaluation of DAE.

\section{The Dataset Used} \label{sec:Dataset}


Aenrouts et al.~\cite{Sigfox_Dataset}  have made publicly available 3 fingerprinting datasets of Low Power Wide Area Networks. Two of these datasets were collected in the urban area of Antwerp, one using Sigfox and another using LoRaWAN. The third dataset was collected in the rural area between the towns of Antwerp and Ghent, using Sigfox. 
Since the initial publication, two major updates have been published. 
In this work, the latest version (1.3) of the LoRaWAN dataset has been utilized, which contains 130430 LoRaWAN messages in the urban area of Antwerp. The fingerprints were collected in an area of approximately 53 square kilometers, though the majority of them lay in the central area of Antwerp which is approximately half the size.

Since the initial publication~\cite{Sigfox_Dataset}, two major updates have been published. 
The fist version (version 1.1) of the LoRaWAN dataset, had a major limitation, as the amount of receiving gateways that reported measurements per message was limited to 3. 
In version 1.2 of the dataset, the urban LoRaWAN dataset has been enriched in terms of new data (130430 LoRaWAN messages), and with measurements from all receiving gateways being reported in the dataset. Moreover, since version 1.2, timestamps have been reported from a number of gateways which are able to provide nanosecond precision timestamps, and which could be used for Time Difference of Arrival (TDOA) localization methods. Lastly, in the most recent version 1.3, the location information of where the gateways of dataset are based is added, a fact that enables the use of the dataset for ranging techniques.



Fingerprinting techniques are often compared to their counterpart, the ranging techniques such as multilateration, which require a minimum of 3 receiving gateways to produce a unique position estimate. Even though satisfying results can be obtained with fingerprinting methods when using messages with fewer than 3 receiving gateways, in this work we limit the dataset to the messages containing at least 3 receiving gateways. A total of 75054 messages with fewer than 3 receiving gateways are dropped, while 55375 messages with at least 3 receiving gateways are kept to be used.
 A common train, validation, test set split will be used for the experiments of this work. We will use 70\% of the dataset for training purposes, 15\% for validation, and 15\% as a test set for reporting the models' performances. In the following sections we will elaborate on the way the training set is repartitioned between two distinct training tasks: the production of position estimates and the production of dynamic accuracy estimates.

\section{Methodology and Terminology} \label{sec:Methodology}  

In the setting of the current work, each message reception obtained by the network of LoRa gateways will be utilized to produce a position estimate, accompanied with a degree of certainty that the system assigns to the position estimate it produced. In this work, this second estimate will be referred to as the Dynamic Accuracy Estimation (DAE), since it concerns an estimation over the accuracy of produced location estimates, which is produced dynamically, as the system operates. The relevant literature refers to the Dynamic Accuracy Estimation either as an `accuracy estimation'~\cite{Moghtadaiee_Accuracy_indicator_2012, Zou_Estimation_2014, Nikitin_laousias_2017}
an `error estimation'~\cite{Dearman_exploration_2007, Lemelson_2009,Marcus_SMARTPOS_2013, Lemic_regression_based_2019,Li_Toward_2019}, or even as a `confidence'~\cite{Elbakly_CONE_2016}. The actual, accurate measurement of the error/accuracy of a position estimate can only become available when the ground truth location is known and its distance from the location estimate produced by the positioning system is measured. 

For the positioning system to produce this twofold estimation, two data-driven models are trained. The first model, $M_1$, is trained to predict the locations from which the message transmissions were made by the mobile device, based on the signals received by the LoRaWAN gateways. In training time, the received signals (in this case the RSS values received at each gateway) are the input features fed into the model, while the target variables are the latitude and longitude of the ground truth location $Loc_{GT}$.
It should be noted that in the dataset used in the current work, the ground truth location is provided by a GPS receiver placed adjacent to the mobile device.

In testing time, a set of previously unseen message receptions are fed into the trained model $M_1$ which produces location estimates $Loc_{Est}$. The error of each produced estimate is defined as the geographic distance $D_{haversine}$ between $Loc_{GT}$ and $Loc_{est}$. In this work, the Haversine distance formula has been used to define this distance.

\begin{equation} \label{eq:Error_M1} 
        Error_{pos} =  Error_{M1} =   D_{haversine}(Loc_{GT},Loc_{est})
\end{equation}

Clarifying the terminology used, we note that Equation~\ref{eq:Error_M1} indicates that the positioning error $Error_{pos}$ of the system corresponds to the prediction error  $Error_{M1}$ of the model $M_1$. This error is measured as the geographic distance $D_{haversine}$ between the ground truth location $Loc_{GT}$ and the estimated location $Loc_{est}$.
The predictive performance of model $M_1$ is evaluated by calculating the errors of all produced location estimates, as by Equation~\ref{eq:Error_M1}, and drawing statistics such as the mean, the median or specific percentile values of the error distribution.

The second model, $M_2$, aims to predict the positioning error $Error_{pos}$ that the location estimate $Loc_{est}$ produced by $M_1$ will have, with only the signals received by the gateways given as input. Therefore, in training time, $M_2$ utilizes the same input features as $M_1$: the received signals of message transmissions (in this case the RSS values received at each gateway), while this time the target variable is the positioning error $Error_{pos}$ that $M_1$'s location estimates have for the same message receptions. In testing time, $M_2$ produces the Dynamic Accuracy Estimation, or $DAE$, that is a prediction regarding the value of $Error_{pos}$ that $M_1$ can achieve for a certain message reception. 

To clarify the concept, we underline that the calculation of $Error_{pos}$ is only feasible when the ground truth location $Loc_{GT}$ is known, thus in the off-line training phase of a fingerprinting system. On the other hand, during the on-line testing phase, or when the system is actually used in production, the ground truth location $Loc_{GT}$ is what the system is aiming to infer. Thus, in those cases $Error_{pos}$ cannot be calculated, and subsequently only an estimate about it can be produced, which is exactly the role of the $DAE$.
In order to quantitatively characterize and evaluate the prediction of $M_2$, the absolute difference between the target variable $Error_{pos}$ and the predicted value $DAE$ are calculated for each prediction, and cumulative statistical metrics such as the mean or the median are produced. Thus, the following metric is commonly used:
\begin{equation} \label{eq:Error_M2} 
        Error_{DAE}  =  Error_{M2} = \lvert Error_{pos}  - DAE \rvert
\end{equation}




\section{Experimentation and Results} \label{sec:Results}
In this section, the performance of model $M_2$, and more particularly, the repercussion that such a model can have on a positioning system as a whole, is studied.
Firstly, since a finite set of fingerprints will be available for the training of the two models ($M_1$, $M_2$), we experiment with the effect that the proportion of training data which are assigned to each of the two models has on their performance (Subsection~\ref{sec:training_set_M1_M2}).
Following, in Subsection~\ref{sec:estimate_selection}, we exemplify the models' capabilities in a use-case where a percentage of the most trustworthy position estimates are selected to be used. Lastly, in Subsection~\ref{sec:Location_estimate_selection_clusters}, we  analyze the performance of $M_2$ in local spatial areas.

Throughout the experiments of this work, the ExtraTrees regression method has been used for the training of both $M_1$ and $M_2$ models. Based on the results of~\cite{Anagnostopoulos2019LoRa} and on preliminary tests performed for this work, the ExtraTrees is preferred to the kNN used in other relevant works. Nevertheless, a detailed comparison among regression methods, which was presented in detail in~\cite{Lemic_regression_based_2019,ANN_Lemic_2020}, is out of the scope of this work. 


\subsection{Training set repartition between $M_1$ and $M_2$} \label{sec:training_set_M1_M2}

\begin{figure}[!h]
  \centering
    \includegraphics[width=0.9\linewidth]{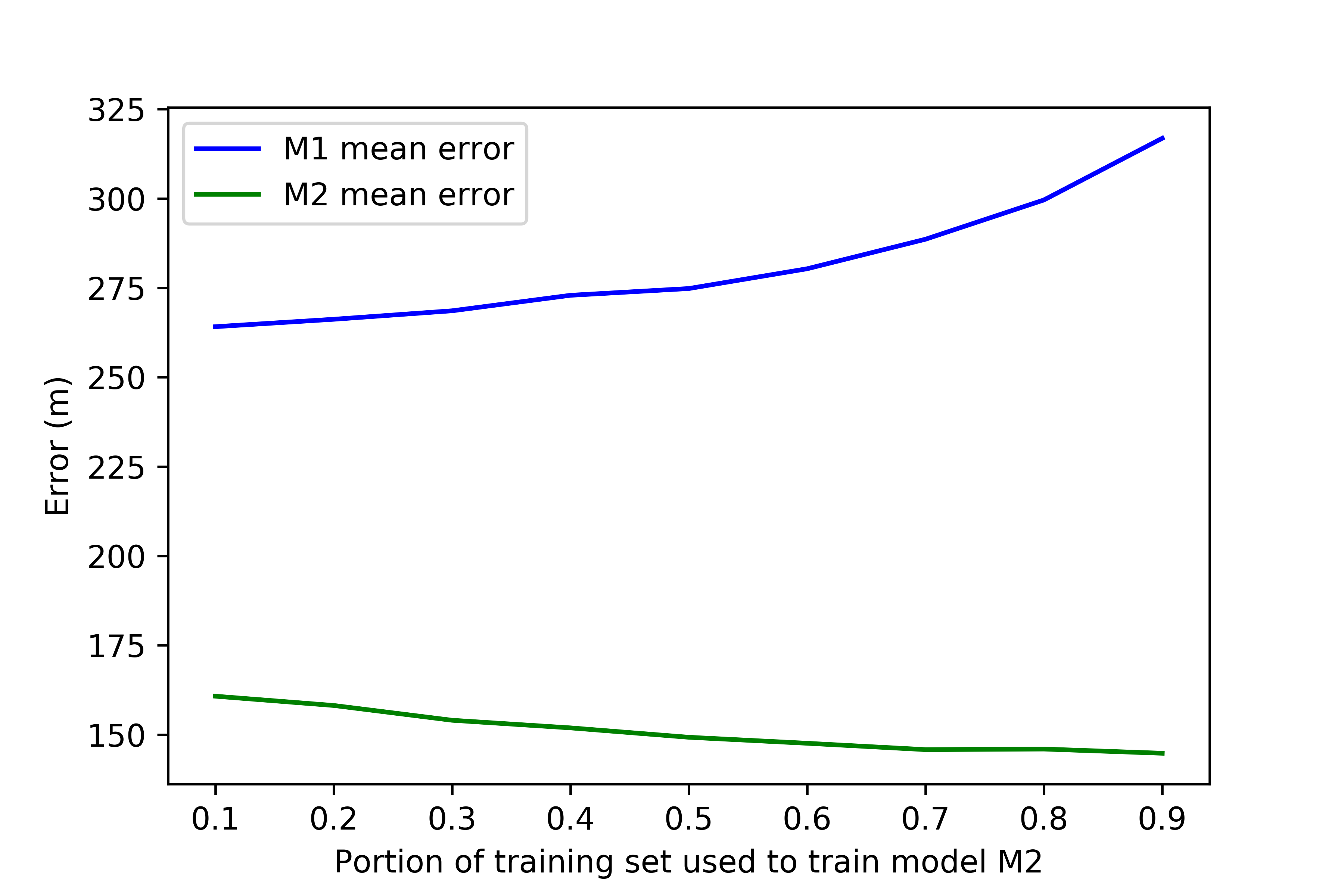}
  \caption{Performance of models $M_1$ and $M_2$ on the validation set, for various portions of the training set assigned to the training of $M_2$. The performance of $M_1$ and $M_2$ corresponds to the mean values of the errors described in Equations \ref{eq:Error_M1} and \ref{eq:Error_M2} respectively, calculated on the validation set.}
  \label{fig:DAE_error_ratios_M1_M2}
\end{figure}

In this first test, we study the performance of the two models for different repartitions of the training set between the two training tasks. The overall dataset has been divided into 70\%, 15\%, 15\% for train, validation and test purposes respectively. At this point, the further repartition of the training set into the two training tasks is studied. The positioning task, served by model $M_1$, is the principal task of the positioning system, while the production of a DAE by a model $M_2$ is naturally a secondary task. Given this fact, it is reasonable for a designer of such a system to assign more data to the training of $M_1$. On the other hand, one may consider particular use-cases in which it might be requested to prioritize a very high certainty regarding the quality of position estimates over the average overall performance of position estimate production. The overview of the performance trade-off between $M_1$ and $M_2$ for different repartitions of the training data is a suitable way of facilitating the designer of such systems to take informed decisions.
Figure \ref{fig:DAE_error_ratios_M1_M2} presents the performance of $M_1$ and $M_2$ on the validation set, for different portions of the training set assigned to the secondary task, dealt by $M_2$.

In Figure \ref{fig:DAE_error_ratios_M1_M2}, it can be observed that the repercussion of the volume of training data is greater on the primary task of positioning dealt by $M_1$  than on the secondary task of DAE dealt by $M_2$.  In model $M_1$, the difference in the mean validation error of $M_1$ between the case of using 90\% of the training data and the case of only using 10\% is 53 meters, which constitutes a 20\% increase of the error. On the other hand, regarding $M_2$, the reduction of the mean validation error of $M_2$ between the cases of using 90\% and 10\% of the training data for its training is only 16 meters, or 11\% of relative error increase.

As discussed previously, the selection of the way that the training data will be allocated to the two tasks depends on the requirements of the application. For the rest of the tests of this work we will proceed with 50\%-50\% sharing of training data between $M_1$ and $M_2$, corresponding to 35\%-35\% of the overall dataset. With this repartition, the validation error of $M_1$ has a mean of 275 meters and a median of 171 meters, while $M_2$ has a 149-meter mean error and an 85-meter median error. The respective test set error for $M_1$ has a mean of 276 meters and a median 176 meters, while $M_2$ has a 147-meter mean error and an 83-meter median error.

\subsection{Location estimate selection based on DAE} \label{sec:estimate_selection}

\begin{figure}[!h]
  \centering
    \includegraphics[width=0.9\linewidth]{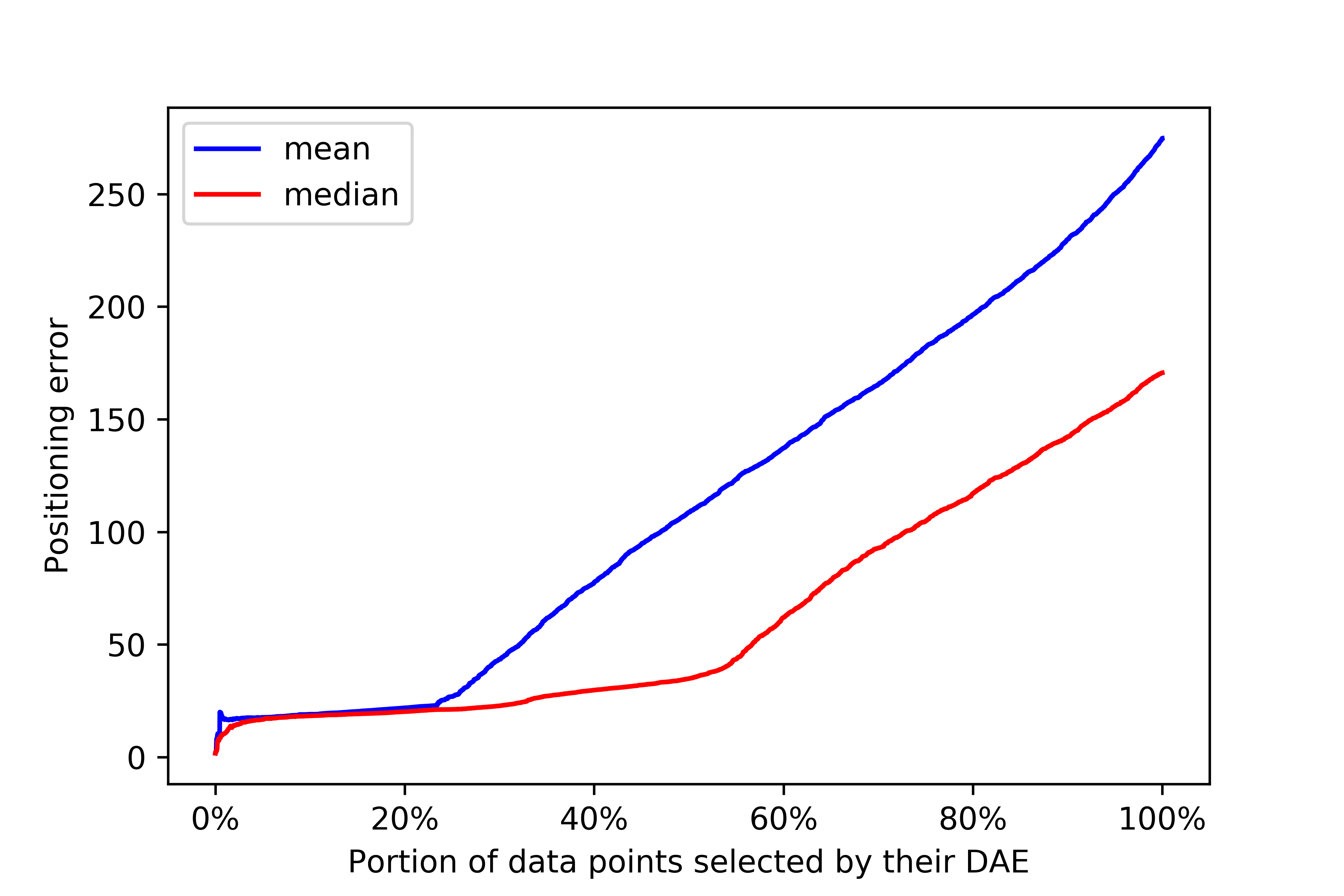}
  \caption{ The mean and median positioning error of portions (subsets) of the validation set, selected according to their DAE value. The data points have been sorted based on the DAE estimated over them.}
  \label{fig:sorted_by_DAE_ml_tril}
\end{figure}

Having trained the two models, we proceed to the exemplification of their combined usage in a process of selecting a subset of position estimates, based on the confidence that the system claims over them. To do so, the following procedure is undertaken.
Initially, both models provide their predictions on the previously unseen validation set: $M_1$ provides the position estimates, and $M_2$ provides the corresponding DAE. The outcomes for all data points of the validation set are sorted with respect to the estimated DAE values, thus starting from the position estimates in which $M_2$ assigns the highest certainty and moving towards the position estimates which $M_2$ predicts as being extremely erroneous. 
Figure~\ref{fig:sorted_by_DAE_ml_tril} offers an overview of the positioning system's performance with respect to different portions of the validation set selected based on the DAE.

Depending on how selective a higher-level system needs to be regarding the accuracy that it requires from the positioning system, the threshold of acceptable DAE values can be set accordingly. The designer of the system may decide regarding the selection method to be used. One approach is to select a certain percentage of the most accurate location estimates, according to the DAE. Alternatively, a hard threshold of the acceptable DAE values can be set, irrespective of the distribution of the DAE values in the studied dataset. 

It is important to note that if it is assumed that the selection process is not done on-line for each message individually as the messages arrive sequentially, but cumulatively upon the collection of a dataset of a certain volume, then the distribution of the DAE values in this set can be considered available. Consequently, in that case, the median or any other percentile value of DAE of such an available set (as for instance, the test set of the current study) can be used as the threshold for selecting the required percentage of position estimates on which $M_2$ has assigned the highest certainty. On the other hand, if the use-case dictates an on-line selection process, the corresponding distribution of the DAE values can be obtained from the validation set, which is known.

The results depicted in Figure~\ref{fig:sorted_by_DAE_ml_tril} suggest that $M_2$ succeeds in estimating the DAE values in a relatively reliable way. In the validation set, in which the positioning model achieves a 275-meter mean error and a 171-meter median error, it is possible to learn to select signal readings of good quality, which produce position estimates whose accuracy is significantly better than the rest. For instance, selecting 50\% of the estimates (those with the lowest estimated error according to DAE), results in a mean positioning error of 108 meters and a median positioning of 35 meters. These values correspond to a 61\% improvement of the mean error and an 80\% improvement of the median error. 

It is noteworthy that, for use-cases such as this one where the goal is the distinction between more accurate and less accurate location estimates, the absolute values of DAE, and consequently its absolute error values described in Equation \ref{eq:Error_M2}, are not as crucial as the relative ordering of the location estimates based on their DAE. In other words, assuming location estimates $i$ and $j$, with $Error_{M1}^{i}<Error_{M1}^{j}$, there are applications where the correct ordering of the $M_2$ estimates, that is $DAE^{i}<DAE^{j}$, may be more important than their absolute error values, $Error_{M2}^{i}$ and $Error_{M2}^{j}$.

\begin{figure}[!h]
\centering
\includegraphics[width=0.99\linewidth]{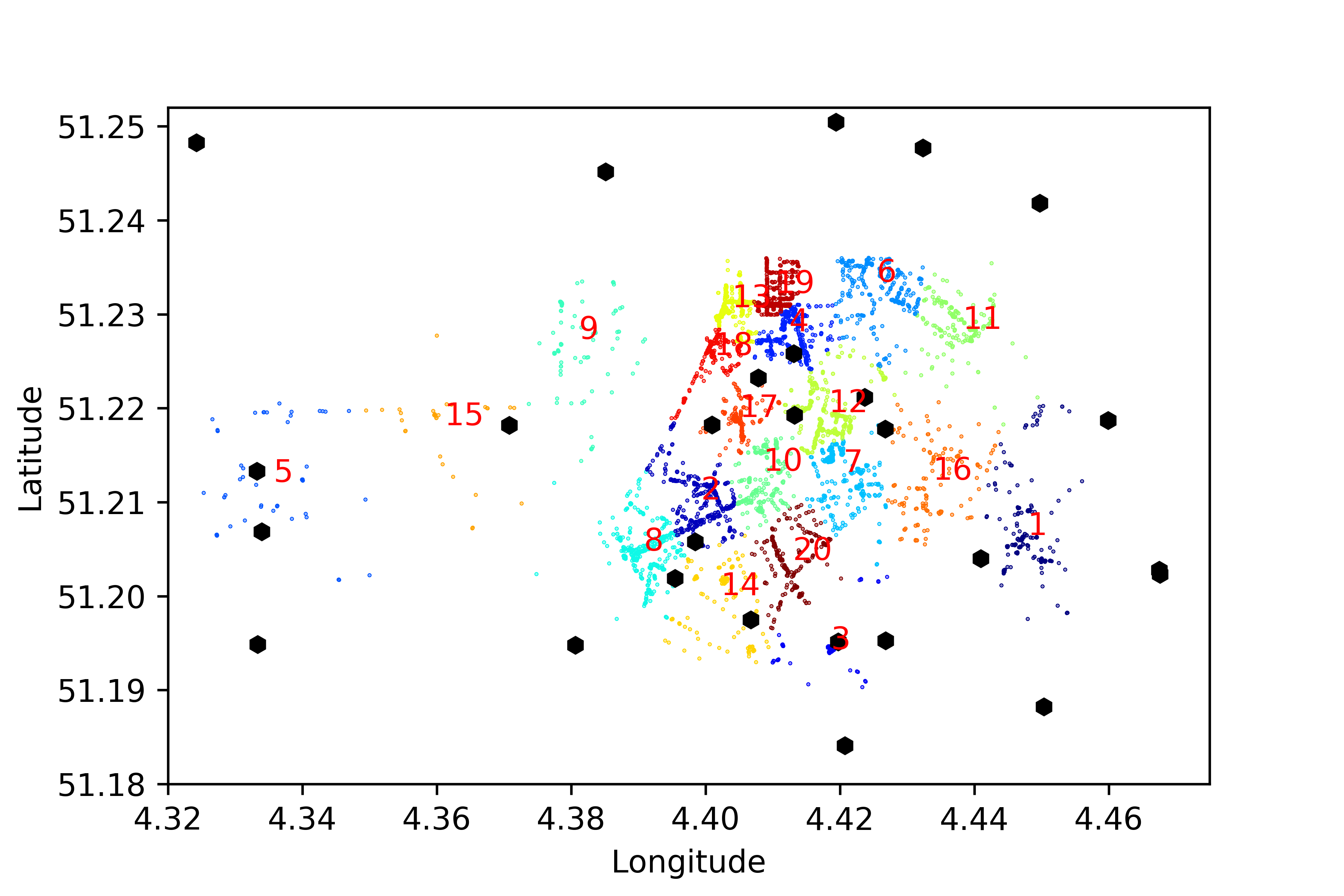}
\caption{The spatial distribution of the data points of the dataset, color coded into 20 clusters. The cluster ID is indicated in red at the cluster centers. Gateway locations are depicted in black.
} \label{fig:kmeans}
\end{figure}

\subsection{Location estimate selection in local clusters} \label{sec:Location_estimate_selection_clusters}

The spatial distribution of the location estimates and its relation to the DAE that is assigned to them by the $M_2$ model is a very interesting subject. The real-world dataset of Antwerp used in this work, like most relevant real-world datasets, does not have a regular, fixed density of collected data throughout different spatial areas. Moreover, using machine learning models implies that their performance is dependent on the quantity of available training data. More specifically, the particular setting of geolocalization further implies that the quantity of data at different areas (thus, the spatial density of training data) determines the models' performance at these areas. This is a consequence of the fact that fingerprinting techniques learn the local particularities of the environment through a mesh of fingerprints. A scarcity of data at a certain zone cannot be compensated for by a profusion of data at a distant area, where most likely a distinct set of gateways will be involved in the signal reception, and thus other features will be of importance.

To study the impact of locality in the two models discussed in this work, the dataset has been clustered into 20 areas, using the k-means algorithm, operating on the ground truth locations of the data. In Figure \ref{fig:kmeans}, the 20 clusters are presented in a color code, having their ID number indicated at the cluster centers. 

It can be observed that the majority of data points has been collected at the central area of Antwerp, at the right-hand side of the map. On the other hand, clusters 5, 9 and 15 are characterized by a rather sparse collection of data. As mentioned previously, the density of data collection at different areas may affect the performance of the positioning system in those areas. Nevertheless, there are other factors that affect the performance of the positioning system as well. The complexity of the environment over which the signals are propagated is a major factor. Factors such as the presence of obstacles (high buildings, hills, etc.) as well as the density and the locations of the gateways may greatly affect a system's performance. For instance, in the city center it is reasonable to expect to have a higher impact of the multipath effect, than in suburban or rural areas. 

\begin{figure}[!h]
  \centering
    \includegraphics[width=0.95\linewidth]{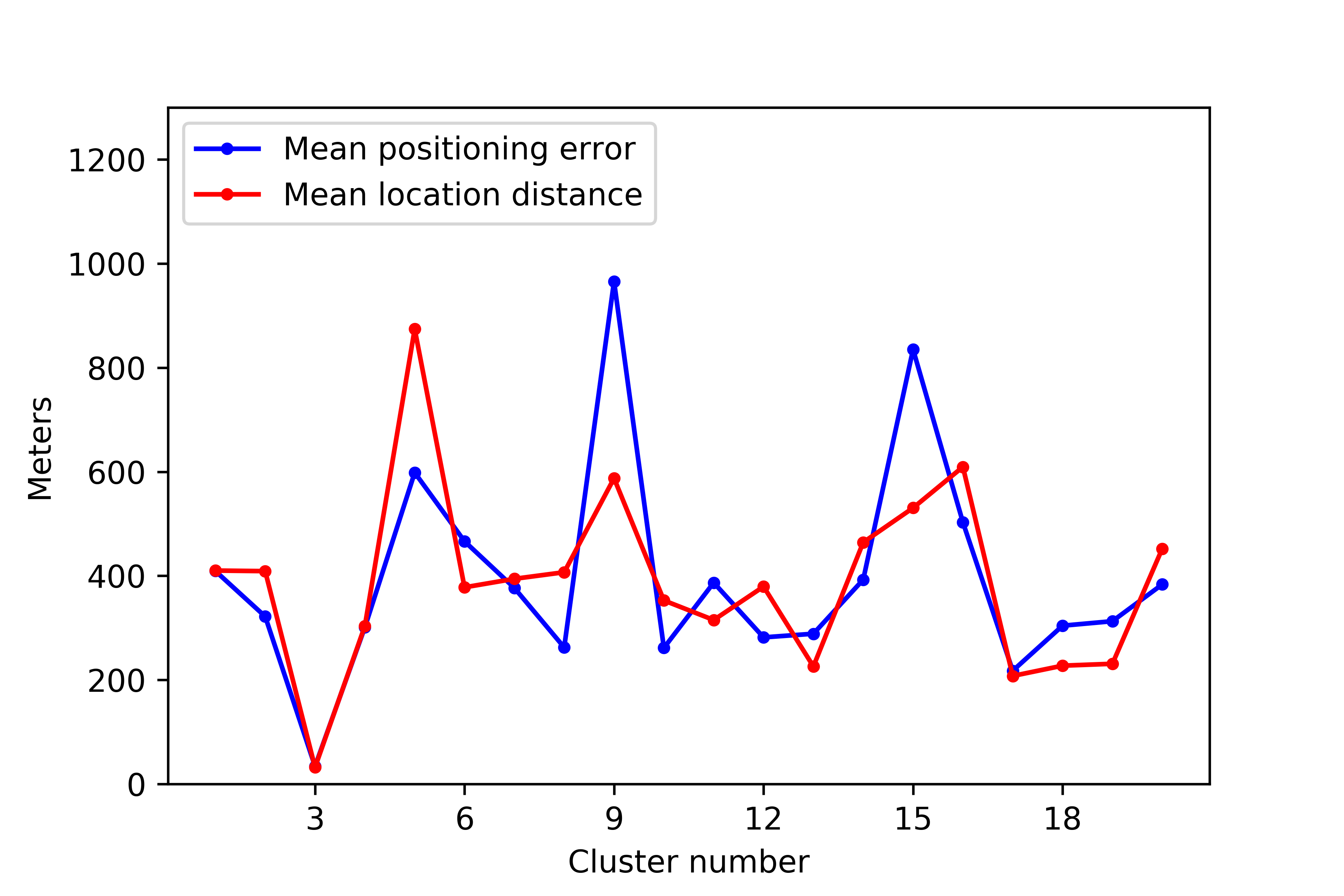}
  \caption{ The $M_1$ error per cluster, along with the mean distance of all location pairs per cluster.}
  \label{fig:M1_error_per_cluster}
\end{figure}

Figure \ref{fig:M1_error_per_cluster} highlights in blue the mean positioning error, for each of the 20 clusters. Furthermore, as a metric indicating the spatial density of data within a cluster, the mean distance of all location pairs in a cluster is indicated in red. The correlation coefficient of the two metrics of Figure~\ref{fig:M1_error_per_cluster} is 0.72. 

\begin{figure}[!h]
  \centering
    \includegraphics[width=0.95\linewidth]{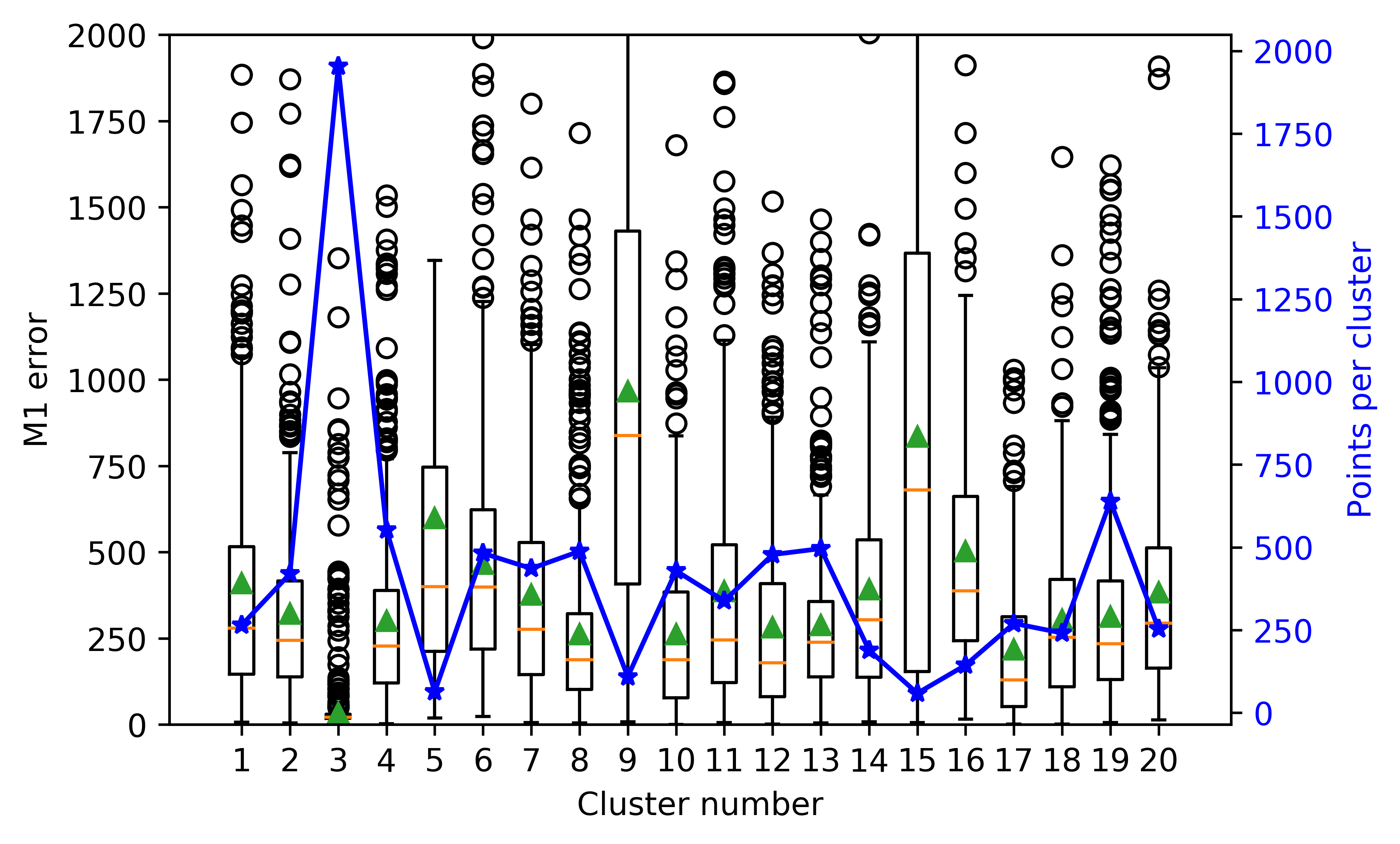}
  \caption{The distribution of the positioning error $M_1$, for each cluster. The mean value is indicated by the green triangle, while in blue the amount of contributing data points per cluster are depicted.}
  \label{fig:boxplots_M1_error_}
\end{figure}

\begin{figure}[!h]
  \centering
    \includegraphics[width=0.95\linewidth]{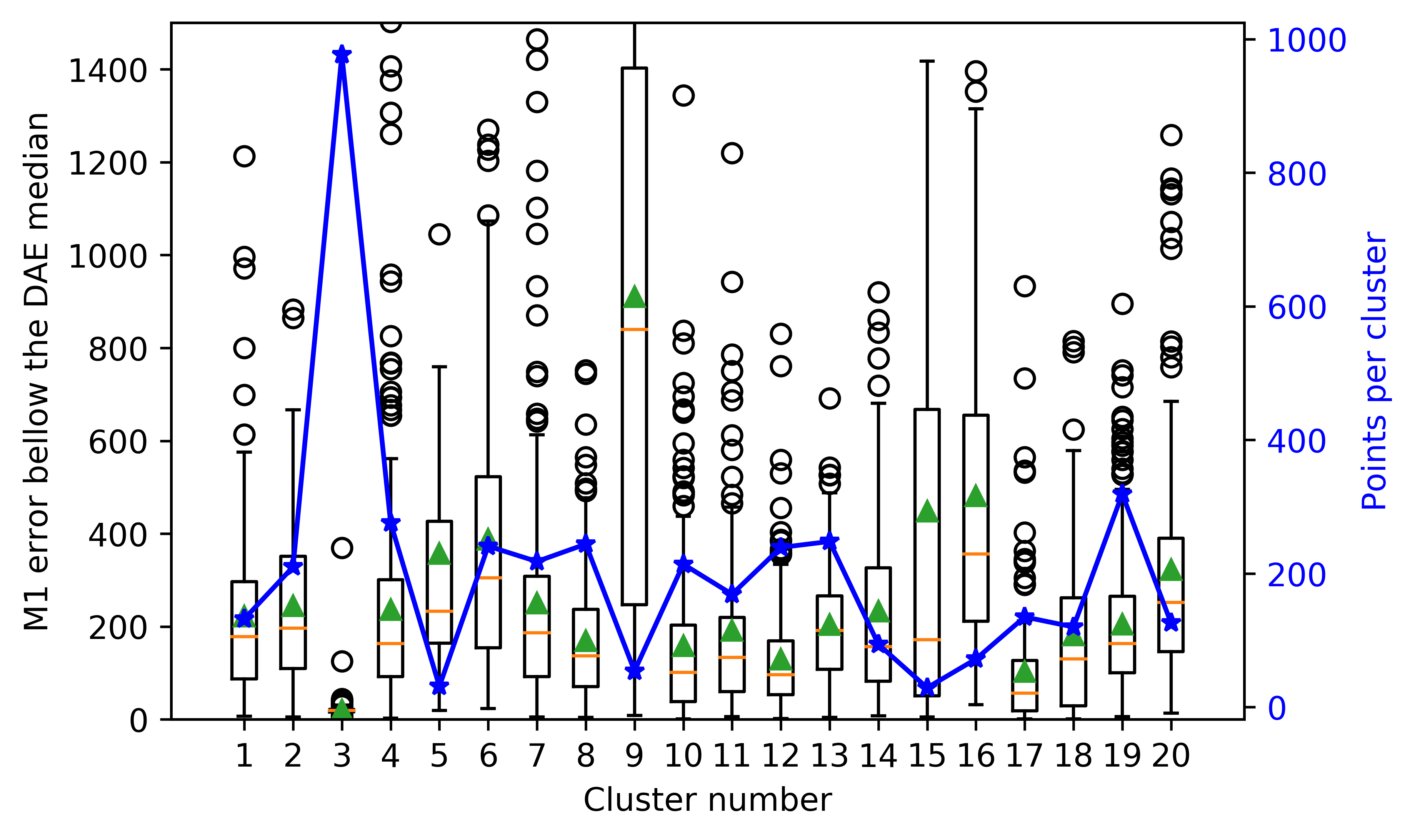}
  \caption{ The distribution of the positioning error $M_1$, for each cluster, for all location estimates that have a bellow median DAE value of their cluster.}
  \label{fig:boxplots_M1_bellow_DAE_median_}
\end{figure}

Figure~\ref{fig:boxplots_M1_error_} presents  the distribution of positioning error of each cluster in the form of boxplots.
Similarly, Figure~\ref{fig:boxplots_M1_bellow_DAE_median_} presents the distribution of positioning error after having selected the 50\% most accurate location estimates of each cluster, based on their DAE, as discussed in Subsection~\ref{sec:estimate_selection}. In both Figures, the number of data points of each cluster contributing to the statistics is depicted in blue. Since the clustering is operated on the ground truth locations of the data, and given the fact that the data have been collected without imposing a requirement of an even density among different areas, the occurring variability in the cluster size emerges unsurprisingly.
The amount of points per cluster has an anticorrelation of coefficient $-0.63$ with the mean positioning error of the clusters in Figure~\ref{fig:boxplots_M1_error_} and of $-0.5$ in Figure~\ref{fig:boxplots_M1_bellow_DAE_median_}. 

These figures provide an insight into how the location estimate selection process may perform when operated locally. 
This is of great interest, since a selection of estimates from the whole dataset may favor the overall error statistics, but under-represent certain regions of high error, and over-represent regions of low error. On the other hand, the evaluation of the selection process performed separately in different areas, approximates in a more realistic way the performance that the selection process would have when performed on-line.
Specifically, when selecting the 50\% of the most accurate (according to DAE) location estimates of each cluster, the mean positioning error of all selected locations is 186 meters and the median 111 meters. The difference is significant when compared to the respective 108-meter mean and 35-meter median error previously reported for the case of selecting the 50\% of the most accurate location estimates from the whole dataset. The significant difference, especially in terms of the median error, is related to the uneven selection of location estimates from different regions.


\section{Conclusions and Future Work}
\label{sec:Conclusions}

In this work, we analyzed various aspects of the data-driven approach of determining the Dynamic Accuracy Estimation (DAE), in the context of its potential use in a broader system. In the previous interesting works that have only recently elaborated on the data-driven approach of DAE determination~\cite{Lemic_regression_based_2019,ANN_Lemic_2020}, the authors have utilized the great majority of the training set (87.6\%) for training the DAE model. Though this option allows the presentation of the performance capabilities of the data-driven determination of the DAE in its full extent, we considered it imperative to investigate the impact that the training data repartition between the two training tasks has on the overall system's performance. In the general case, it is natural to prioritize the positioning task as the primary task of the system, assigning to it more training data, rather than the DAE determination task. Moreover, the experimentation of the current work indicated that when receiving more training data, the positioning model has a considerable gain in terms of error reduction (both in absolute and percentage terms), than what the DAE determining model has in the same case. 
It is up to the designer of a system to choose the repartition that fits the use-case that the system serves.
This work has exemplified, in Subsection~\ref{sec:training_set_M1_M2}, the way to obtain an overview of this trade-off, which allows informed decisions to be taken.

In Subsection~\ref{sec:estimate_selection}, we have elaborated on the fact that DAE, apart from its evident usefulness as an additional information to the user, can have other interesting applications when utilized by the overall system. A utilization of DAE for the selection of the most accurate location estimates has been exemplified, returning to the user a subset of estimates with a significantly increased location accuracy. Such a process may be used in use-cases where such an option would be appropriate.

Moreover, in Subsection~\ref{sec:Location_estimate_selection_clusters} we studied the performance of the two models in local, spatial areas via the use of clustering. In the particular setting under study, it was observed that certain areas (cluster 3) may have a high concentration of data, yielding very accurate predictions from both models, in an unbalanced manner compared to the rest of the dataset. Secondly, it was observed that the suburban areas (clusters 5, 9 and 15) had the highest positioning error. These three clusters were also among the most sparsely collected areas. 
Overall, we identified that the mean positioning error per cluster had a significant correlation with the average distance among (or the density of) the data points per cluster, and a significant anticorrelation with the amount of data points per cluster. 
 Lastly, we have highlighted the fact that in datasets with such diverse density of data collected among different spatial zones, it is interesting to look into performance statistics from spatial zones along with the statistics of the overall dataset.

In the spirit of repeatability, reproducibility, verifiability and comparability of results, it is important not only to report performance metrics over a public dataset, but to also provide the means for reproducing the same train/validation/test set split. Ideally, the code, which unambiguously presents the exact way that the reported results were produced, may be published as well. In this way, the community is able to repeat, reproduce, verify and consistently compare results, and is eventually enabled to take informed decisions over the way positioning systems are designed, implemented and deployed. For these reasons, the full code implementation and the train/validation/test sets of the current work are openly available at the Zenodo repository.



%
\balance

\bibliographystyle{IEEEtran}
\bibliography{bibliography}

\begin{thebibliography}{10}
\providecommand{\url}[1]{#1}
\csname url@samestyle\endcsname
\providecommand{\newblock}{\relax}
\providecommand{\bibinfo}[2]{#2}
\providecommand{\BIBentrySTDinterwordspacing}{\spaceskip=0pt\relax}
\providecommand{\BIBentryALTinterwordstretchfactor}{4}
\providecommand{\BIBentryALTinterwordspacing}{\spaceskip=\fontdimen2\font plus
\BIBentryALTinterwordstretchfactor\fontdimen3\font minus
  \fontdimen4\font\relax}
\providecommand{\BIBforeignlanguage}[2]{{%
\expandafter\ifx\csname l@#1\endcsname\relax
\typeout{** WARNING: IEEEtran.bst: No hyphenation pattern has been}%
\typeout{** loaded for the language `#1'. Using the pattern for}%
\typeout{** the default language instead.}%
\else
\language=\csname l@#1\endcsname
\fi
#2}}
\providecommand{\BIBdecl}{\relax}
\BIBdecl

\bibitem{Sigfox_Dataset}
\BIBentryALTinterwordspacing
M.~Aernouts, R.~Berkvens, K.~Van~Vlaenderen, and M.~Weyn, ``Sigfox and lorawan
  datasets for fingerprint localization in large urban and rural areas,''
  \emph{Data}, vol.~3, no.~2, 2018. [Online]. Available:
  \url{http://www.mdpi.com/2306-5729/3/2/13}
\BIBentrySTDinterwordspacing

\bibitem{Janssen_Sigfox}
T.~{Janssen}, M.~{Aernouts}, R.~{Berkvens}, and M.~{Weyn}, ``Outdoor
  fingerprinting localization using sigfox,'' in \emph{2018 International
  Conference on Indoor Positioning and Indoor Navigation (IPIN)}, Sep. 2018,
  pp. 1--6.

\bibitem{Anagnostopoulos2019LoRa}
G.~G. {Anagnostopoulos} and A.~{Kalousis}, ``A reproducible comparison of rssi
  fingerprinting localization methods using lorawan,'' in \emph{2019 16th
  Workshop on Positioning, Navigation and Communications (WPNC)}, 2019, pp.
  1--6.

\bibitem{Janssen_2020_Benchmarking}
\BIBentryALTinterwordspacing
T.~Janssen, R.~Berkvens, and M.~Weyn, ``Benchmarking rss-based localization
  algorithms with lorawan,'' \emph{Internet of Things}, vol.~11, p. 100235,
  2020. [Online]. Available:
  \url{http://www.sciencedirect.com/science/article/pii/S2542660520300688}
\BIBentrySTDinterwordspacing

\bibitem{Anagnostopoulos2019_IPIN}
G.~G. {Anagnostopoulos} and A.~{Kalousis}, ``A reproducible analysis of rssi
  fingerprinting for outdoor localization using sigfox: Preprocessing and
  hyperparameter tuning,'' in \emph{2019 International Conference on Indoor
  Positioning and Indoor Navigation (IPIN)}, 2019, pp. 1--8.

\bibitem{Hybrid_LoRaWAN_2021}
Z.~A. {Pandangan} and M.~C.~R. {Talampas}, ``Hybrid lorawan localization using
  ensemble learning,'' in \emph{2020 Global Internet of Things Summit (GIoTS)},
  2020, pp. 1--6.

\bibitem{Lemic_regression_based_2019}
F.~{Lemic}, V.~{Handziski}, M.~{Aernouts}, T.~{Janssen}, R.~{Berkvens},
  A.~{Wolisz}, and J.~{Famaey}, ``Regression-based estimation of individual
  errors in fingerprinting localization,'' \emph{IEEE Access}, vol.~7, pp.
  33\,652--33\,664, 2019.

\bibitem{ANN_Lemic_2020}
F.~{Lemic} and J.~{Famaey}, ``Artificial neural network-based estimation of
  individual localization errors in fingerprinting,'' in \emph{2020 IEEE 17th
  Annual Consumer Communications Networking Conference (CCNC)}, 2020, pp. 1--6.

\bibitem{Torres_Sospedra_competition_2017}
\BIBentryALTinterwordspacing
J.~Torres-Sospedra, A.~Jiménez, A.~Moreira, T.~Lungenstrass, W.-C. Lu,
  S.~Knauth, G.~Mendoza-Silva, F.~Seco, A.~Pérez-Navarro, M.~Nicolau, and
  et~al., ``Off-line evaluation of mobile-centric indoor positioning systems:
  The experiences from the 2017 ipin competition,'' \emph{Sensors}, vol.~18,
  no.~2, p. 487, Feb 2018. [Online]. Available:
  \url{http://dx.doi.org/10.3390/s18020487}
\BIBentrySTDinterwordspacing

\bibitem{GpmLab_2016}
C.~M. {de la Osa}, G.~G. {Anagnostopoulos}, M.~{Togneri}, M.~{Deriaz}, and
  D.~{Konstantas}, ``Positioning evaluation and ground truth definition for
  real life use cases,'' in \emph{2016 International Conference on Indoor
  Positioning and Indoor Navigation (IPIN)}, Oct 2016, pp. 1--7.

\bibitem{GpmStudio_2016}
G.~G. {Anagnostopoulos}, C.~M. {de la Osa}, T.~{Nunes}, A.~{Hammoud},
  M.~{Deriaz}, and D.~{Konstantas}, ``Practical evaluation and tuning
  methodology for indoor positioning systems,'' in \emph{2016 Fourth
  International Conference on Ubiquitous Positioning, Indoor Navigation and
  Location Based Services (UPINLBS)}, 2016, pp. 130--139.

\bibitem{GpmStudio_MO_2017}
G.~G. {Anagnostopoulos}, M.~{Deriaz}, and D.~{Konstantas}, ``A multiobjective
  optimization methodology of tuning indoor positioning systems,'' in
  \emph{2017 International Conference on Indoor Positioning and Indoor
  Navigation (IPIN)}, 2017, pp. 1--8.

\bibitem{Lemelson_2009}
H.~Lemelson, M.~B. Kj{\ae}rgaard, R.~Hansen, and T.~King, ``Error estimation
  for indoor 802.11 location fingerprinting,'' in \emph{Location and Context
  Awareness}, T.~Choudhury, A.~Quigley, T.~Strang, and K.~Suginuma, Eds.\hskip
  1em plus 0.5em minus 0.4em\relax Berlin, Heidelberg: Springer Berlin
  Heidelberg, 2009, pp. 138--155.

\bibitem{Marcus_SMARTPOS_2013}
P.~Marcus, M.~Kessel, and M.~Werner, ``Dynamic nearest neighbors and online
  error estimation for smartpos,'' \emph{International Journal On Advances in
  Internet Technology}, vol.~6, pp. 1--11, 01 2013.

\bibitem{Li_Toward_2019}
Y.~{Li}, Z.~{He}, Z.~{Gao}, Y.~{Zhuang}, C.~{Shi}, and N.~{El-Sheimy}, ``Toward
  robust crowdsourcing-based localization: A fingerprinting accuracy indicator
  enhanced wireless/magnetic/inertial integration approach,'' \emph{IEEE
  Internet of Things Journal}, vol.~6, no.~2, pp. 3585--3600, April 2019.

\bibitem{Moghtadaiee_Accuracy_indicator_2012}
V.~{Moghtadaiee}, A.~G. {Dempster}, and B.~{Li}, ``Accuracy indicator for
  fingerprinting localization systems,'' in \emph{Proceedings of the 2012
  IEEE/ION Position, Location and Navigation Symposium}, 2012, pp. 1204--1208.

\bibitem{Zou_Estimation_2014}
D.~{Zou}, W.~{Meng}, and S.~{Han}, ``An accuracy estimation algorithm for
  fingerprint positioning system,'' in \emph{2014 Fourth International
  Conference on Instrumentation and Measurement, Computer, Communication and
  Control}, 2014, pp. 573--577.

\bibitem{Elbakly_CONE_2016}
\BIBentryALTinterwordspacing
R.~Elbakly and M.~Youssef, ``{CONE:} zero-calibration accurate confidence
  estimation for indoor localization systems,'' \emph{CoRR}, vol.
  abs/1610.02274, 2016. [Online]. Available:
  \url{http://arxiv.org/abs/1610.02274}
\BIBentrySTDinterwordspacing

\bibitem{Syed_Khandker_2019}
S.~{Khandker}, R.~{Mondal}, and T.~{Ristaniemi}, ``Positioning error prediction
  and training data evaluation in rf fingerprinting method,'' in \emph{2019
  International Conference on Indoor Positioning and Indoor Navigation (IPIN)},
  2019, pp. 1--7.

\bibitem{Nikitin_laousias_2017}
A.~{Nikitin}, C.~{Laoudias}, G.~{Chatzimilioudis}, P.~{Karras}, and
  D.~{Zeinalipour-Yazti}, ``Indoor localization accuracy estimation from
  fingerprint data,'' in \emph{2017 18th IEEE International Conference on
  Mobile Data Management (MDM)}, 2017, pp. 196--205.

\bibitem{Energy_Accuracy_Lin_2010}
K.~Lin, A.~Kansal, D.~Lymberopoulos, and F.~Zhao, ``Energy-accuracy trade-off
  for continuous mobile device location,'' in \emph{Proceedings of the 8th
  International Conference on Mobile Systems, Applications, and Services}, ser.
  MobiSys '10.\hskip 1em plus 0.5em minus 0.4em\relax Association for Computing
  Machinery, 2010, p. 285–298.

\bibitem{Zou_handoff_2013}
{Deyue Zou}, {Weixiao Meng}, and {Shuai Han}, ``Euclidean distance based
  handoff algorithm for fingerprint positioning of wlan system,'' in \emph{2013
  IEEE Wireless Communications and Networking Conference (WCNC)}, 2013, pp.
  1564--1568.

\bibitem{Anagnostopoulos_IPIN_2015}
G.~G. {Anagnostopoulos} and M.~{Deriaz}, ``Automatic switching between indoor
  and outdoor position providers,'' in \emph{2015 International Conference on
  Indoor Positioning and Indoor Navigation (IPIN)}, 2015, pp. 1--6.

\bibitem{Dearman_exploration_2007}
D.~Dearman, A.~Varshavsky, E.~de~Lara, and K.~N. Truong, ``An exploration of
  location error estimation,'' in \emph{UbiComp 2007: Ubiquitous Computing},
  J.~Krumm, G.~D. Abowd, A.~Seneviratne, and T.~Strang, Eds.\hskip 1em plus
  0.5em minus 0.4em\relax Berlin, Heidelberg: Springer Berlin Heidelberg, 2007,
  pp. 181--198.

\end{thebibliography}

\end{document}